\documentclass[final,a4paper,UKenglish,cleveref,autoref]{oasics-v2021}

\title{A Formal Approach to AMM Fee Mechanisms with {Lean} 4}

\author{Marco Dessalvi}{Technical University of Denmark, Denmark}{m.dessalvi02@gmail.com}{}{}

\author{Massimo Bartoletti}{University of Cagliari, Italy \and \url{http://blockchain.unica.it} }{bart@unica.it}{https://orcid.org/0000-0003-3796-9774}{Partially supported by project SERICS (PE00000014) and PRIN 2022 DeLiCE (F53D23009130001) under the MUR National Recovery and Resilience Plan funded by the European Union -- NextGenerationEU.}

\author{Alberto Lluch-Lafuente}{Technical University of Denmark, Denmark}{albl@dtu.dk}{https://orcid.org/0000-0001-7405-0818}{Partially supported by Copenhagen Fintech project CODeM - Challenges and Opportunities of Defi Models, special thanks to Troels Damgaard, Espen Højsgaard, and Lasse Nisted for fruitful discussions on DeFi fee mechanisms.}

\authorrunning{M. Dessalvi, M. Bartoletti, A. Lluch-Lafuente} 

\Copyright{M. Dessalvi, M. Bartoletti, A. Lluch-Lafuente}



\nolinenumbers 

\usepackage[utf8]{inputenc}

\usepackage{tikz}           
\usepackage{pgfplots}       
\usepackage{xcolor}         
\usepackage{amsmath}        
\usepackage{siunitx}        
\usepackage{listings}       
\PassOptionsToPackage{hyphens}{url} 
\usepackage{hyperref}       
\usepackage{cleveref}       
\usepackage{textcomp}       

\usepackage{caption}        
\usepackage{subcaption}     
\usepackage[utf8]{inputenc}
\usepackage{csquotes}       
\usepackage{tabularx}       
\usepackage{booktabs}       
\usepackage{float}          
\usepackage{calc}           
\usepackage{blindtext}

\usepackage{color}
\usepackage{xspace} 
\usepackage{bm} 
\usepackage{amssymb}
\usepackage{amsthm}
\usepackage{graphicx}
\usepackage{mathtools}
\usepackage{url}
\usepackage{etoolbox}
\usepackage[inline,shortlabels]{enumitem} 
\newlist{inlinelist}{enumerate*}{1}
\setlist*[inlinelist,1]{%
  label=(\roman*),
}
\usepackage{xifthen}  
\usepackage{ifthen}
\usepackage{nicefrac}
\usepackage{hyperref}
\usepackage{cleveref}
\hypersetup{
  breaklinks   = true,
  colorlinks   = true, 
  urlcolor     = blue, 
  linkcolor    = blue, 
  citecolor    = red   
}
\hypersetup{final} 
\usepackage[draft,nomargin,inline,index]{fixme} 

\fxusetheme{color}

\definecolor{BlueViolet}{rgb}{0, 0, 0.55}
\definecolor{RubineRed}{rgb}{0.88, 0.07, 0.37}
\definecolor{ForestGreen}{rgb}{0.13, 0.55, 0.13}
\definecolor{Blue}{rgb}{0.0, 0.0, 1.0}
\definecolor{NavyBlue}{rgb}{0.0, 0.0, 0.5}
\definecolor{Black}{rgb}{0.02, 0.02, 0.02}
\definecolor{MidnightBlue}{rgb}{0.0, 0.2, 0.4}
\definecolor{Gray}{rgb}{0.41, 0.41, 0.41}
\definecolor{TealBlue}{rgb}{0.212,0.459,0.533}
\definecolor{Plum}{rgb}{0.6,0.25,0.6}

\definecolor{Black}{HTML}{000000}
\definecolor{Gray}{HTML}{808080}
\definecolor{Magenta}{HTML}{FF00FF}
\definecolor{RubineRed}{HTML}{ED017D}
\definecolor{ForestGreen}{HTML}{028A0F}
\definecolor{OliveGreen}{HTML}{808000}
\definecolor{MidnightBlue}{HTML}{006795}
\definecolor{Plum}{HTML}{92268F}

\FXRegisterAuthor{bart}{anbart}{\color{magenta} {\underline{bart}}}
\FXRegisterAuthor{marco}{anmarco}{\color{red} {\underline{marco}}}
\FXRegisterAuthor{alb}{analberto}{\color{blue} {\underline{alberto}}}

\usepackage{tcolorbox}
\tcbuselibrary{theorems, skins, breakable}

\allowdisplaybreaks
\usepackage{amsmath,amsthm}
\usepackage[framemethod=tikz]{mdframed}
\usepackage{etoolbox}
\usepackage{mdframed}

\newtcbtheorem{defi}{Definition}%
  {colback=green!3, 
  colframe=green!60!black,
  fonttitle=\bfseries,
  before skip=12pt, after skip=12pt}{def}

\newtcbtheorem{theoremi}{Theorem}%
  {colback=blue!10, 
  colframe=blue!75!black,
  fonttitle=\bfseries,
  before skip=12pt, after skip=12pt,
  separator sign={\ $\blacktriangleright$}}{thm}

\newcommand{\ifempty}[3]{%
  \ifthenelse{\isempty{#1}}{#2}{#3}%
}

\newcommand{\ifdots}[3]{%
  \ifthenelse{\equal{#1}{...}}{#2}{#3}%
}

\newcommand{\hidden}[1]{}

\newcommand{\keyterm}[1]{\textbf{\emph{#1}}}%

\newcommand{\Real}[1]{\mathrm{Real}}

\newcommand{\eg}{e.g.\@\xspace}
\newcommand{\ie}{i.e.\@\xspace}
\newcommand{\wrt}{w.r.t.\@\xspace}

\renewcommand{\epsilon}{\varepsilon}

  {\myqed}

\newcommand{\BTC}{\textup{%
  \leavevmode
  \vtop{\offinterlineskip 
    \setbox0=\hbox{B}%
    \setbox2=\hbox to\wd0{\hfil\hskip-.03em
    \vrule height .3ex width .15ex\hskip .08em
    \vrule height .3ex width .15ex\hfil}
    \vbox{\copy2\box0}\box2}}\xspace}

\def\pmvColor{\color{ForestGreen}}

\newcommand{\pmvFmt}[1]{{\pmvColor{\sf{#1}}}} 

\newcommand{\pmv}[2][]{\pmvFmt{#2}_{\pmvColor{#1}}\xspace}

\newcommand{\pmvA}[1][]{\pmv[{#1}]{A}}

\newcommand{\pmvB}[1][]{\pmv[{#1}]{B}}

\newcommand{\supply}[2][]{\mathit{S}_{#1}{#2}} 

\newcommand{\gain}[3][]{\mathit{G}_{#2}(\ifempty{#1}{}{{#1},}{#3})}

\def\txColor{\color{MidnightBlue}}

\newcommand{\txFmt}[1]{{\txColor{\sf #1}}}

\newcommand{\tx}[2][]{\txFmt{#2}_{\txColor{#1}}}
\newcommand{\txT}[1][]{\tx[#1]{T}}

\DeclareMathAlphabet{\mathbfsf}{\encodingdefault}{\sfdefault}{bx}{n}

\newcommand{\nrule}[1]{{\scriptsize \textsc{#1}}}

\newcommand{\setenum}[1]{\{#1\}}

\newcommand{\qedex}{\ensuremath{\diamond}}

\crefname{appendix}{appendix}{appendices}
\Crefname{appendix}{Appendix}{Appendices}
\crefname{notation}{notation}{notations}
\Crefname{notation}{Notation}{Notations}

\definecolor{LightGrey}{rgb}{0.95,0.95,0.95}
\definecolor{keyword}{HTML}{7F0055}

\def\tokColor{\color{magenta}}
\newcommand{\tokFmt}[1]{{\tokColor{#1}}}

\newcommand{\tokM}[2]{\setenum{{#1},{#2}}}

\newcommand{\tokT}[1][]{\tokFmt{\tau_{#1}}}
\newcommand{\tokTi}[1][]{\tokFmt{\tau'_{#1}}}

\newcommand{\TokU}[1][]{\tokFmt{\mathbb{T}_{#1}}} 

\newcommand{\ammSwapOp}{{\txColor{\sf swap}}}

\newcommand{\ammSwapParOp}[1]{{\txColor{\sf swap}^{#1}}}

\newcommand{\actAmmSwapExact}[5]{\ifempty{#1}{}{{#1}:}\ammSwapParOp{#2}({#3},{#4},{#5})}

\newcommand{\tokBal}[1][]{\sigma_{#1}} 
\newcommand{\wal}[2]{{#1}[{#2}]}
\newcommand{\walA}[2][]{\wal{\pmvA[#1]}{#2}}

\newcommand{\stdSupply}{\supply[\confG]{\tokM{\tokT[0]}{\tokT[1]}}}
\newcommand{\stdWallMint}{\tokBal[\pmvA]{\tokM{\tokT[0]}{\tokT[1]}}}

\newcommand{\valSXa}{\alpha}

\newcommand{\valSXb}{\beta}

\newcommand{\X}[2][]{\ifempty{#1}{X}{X_{#1}}\ifempty{#2}{}{({#2})}}

\newcommand{\exchO}[2][]{\ifempty{#1}{P}{P_{#1}}\ifempty{#2}{}{({#2})}}

\newcommand{\SX}[2][]{\mathit{SX}_{\phi_{#1}}\ifempty{#2}{}{({#2})}}

\newcommand{\Z}[4]{\zeta_{\fee} (#1, #2, #3, #4)}

\newcommand{\confG}[1][]{\Gamma_{#1}}
\newcommand{\confGi}[1][]{\Gamma'_{#1}}
\newcommand{\confGii}[1][]{\Gamma''_{#1}}

\newcommand{\confD}[1][]{\Delta_{#1}}

\newcommand{\amm}[2]{\setenum{{#1},{#2}}} 
\newcommand{\ammR}[1][]{r_{#1}} 
\newcommand{\ammRi}[1][]{r'_{#1}}

\newcommand{\fee}{\phi} 

\newcommand\numberthis{\addtocounter{equation}{1}\tag{\theequation}}

\newcommand{\Rpos}{\mathbb{R}_{>0}}

\newcommand{\citeLeanRoot}{https://mamboleano.github.io/lean4-amm-fees/}

\newcommand{\citeLeanLink}[1]{\href{\citeLeanRoot/#1}{\includegraphics[scale=0.1]{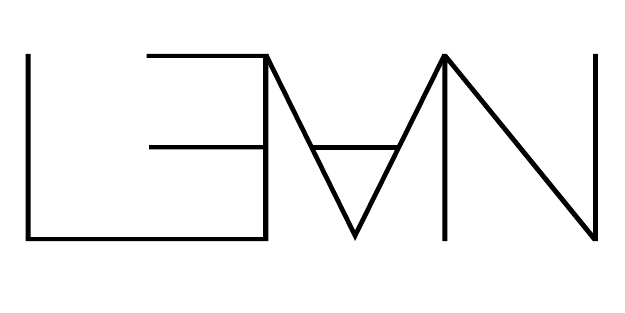}}}

\newcommand{\citeLean}[1]{\vspace{10pt}\llap{\vspace{26pt}\citeLeanLink{#1} \quad\vspace{-55pt}}}

\usepackage{xcolor}
\usepackage{listings}
\usepackage{listings-solidity}   

\definecolor{verylightgray}{rgb}{.97,.97,.97}
\newcommand{\codelst}[1]{\lstinline[basicstyle=\ttfamily\small]!#1!}

\makeatletter
  \renewcommand\footnotesize{\scriptsize}
\makeatother

\begin{document}

\maketitle

\begin{abstract}
Decentralized Finance (DeFi) has revolutionized financial markets by enabling complex asset-exchange protocols without trusted intermediaries. 
Automated Market Makers (AMMs) are a central component of DeFi, providing the core functionality of swapping assets of different types at algorithmically computed exchange rates.
Several mainstream AMM implementations are based on the constant-product model, which ensures that swaps preserve the product of the token reserves in the AMM --- up to a \emph{trading fee} used to incentivize liquidity provision.
Trading fees substantially complicate the economic properties of AMMs, and for this reason some AMM models abstract them away in order to simplify the analysis. 
However, trading fees have a non-trivial impact on users' trading strategies, making it crucial to develop refined AMM models that precisely account for their effects.
In this work, we extend a foundational model of AMMs by introducing a new parameter, the trading fee $\fee \in (0,1]$, into the swap rate function. Fee amounts increase inversely proportional to $\phi$. When $\fee = 1$, no fee is applied and the original model is recovered.
We analyze the resulting fee-adjusted model from an economic perspective.
We show that several key properties of the swap rate function, including output-boundedness and monotonicity, are preserved. 
At the same time, other properties --- most notably additivity --- no longer hold. 
We precisely characterize this deviation by deriving a generalized form of additivity that captures the effect of swaps in the presence of trading fees. 
In particular, we prove that when $\fee < 1$, executing a single large swap yields strictly greater profit than splitting the trade into smaller ones.
Finally, we derive a closed-form solution to the arbitrage problem in the presence of trading fees and prove its uniqueness. 
All results are formalized and machine-checked in the Lean 4 proof assistant.
\end{abstract}


\section{Introduction}
\label{sec:intro}

Decentralized Finance (DeFi) has transformed the financial landscape by enabling peer-to-peer transactions without a central intermediary. One of the core mechanisms of many DeFi applications is the Automated Market Makers (AMMs), which enable token swaps through algorithmic pricing mechanisms. Among the various kinds of AMMs, the constant-product Market Makers (CPMMs) are the most used ones, with implementations such as Uniswap v2/v3 \cite{uniswapv2,uniswapv3}, Sushiswap \cite{sushiswapimpl}, \emph{etc}. These implementations have a huge economic relevance in the crypto market: for example, as of January 2026, Uniswap stands out with an approximate Total Value Locked (TVL) of \$4 billion \cite{mottaghi2025tvl}. 

While AMMs have been extensively studied, the role of \emph{trading fees} --- which is critical in real-world AMM implementations --- has often been overlooked in formal analyses of AMM models~\cite{BCL22lmcs,Pusceddu24fmbc,Renieri25bcra}. 
Trading fees play a central role in incentivizing liquidity provision, while also significantly influencing users' strategies and overall market dynamics. 
A formal understanding of how trading fees impact AMM behavior is essential both from a theoretical standpoint and practical applications such as constructing of optimal automatic trading bots.

\paragraph*{Contributions}

We provide a formal analysis of AMMs incorporating trading fees, extending a Lean 4 library for no-fee AMMs~\cite{Pusceddu24fmbc}.
More specifically, we summarize our main contributions as follows:
\begin{enumerate}

\item \textbf{Formal Modeling}: We develop a formal model of AMMs that integrates trading fees, capturing their effect on swap outcomes and users' incentives. We introduce this trading fee in the \emph{constant-product} swap rate function, and discuss the similarities between this model and the implementation of the swap in Uniswap v2. 
    
\item \textbf{Economic Analysis}: We analyse core economic properties of AMMs  with trading fees, such as how they impact user gain on token swaps, arbitrage opportunities and exchange rates. 
Besides re-proving the structural properties known from the fee-less analysis of~\cite{BCL22lmcs} in our fee-adjusted setting (\Cref{lem:fee:constprod:output-bound,lem:fee:constprod:strict-mono}), we generalise the additivity property of the constant-product swap rate function (\Cref{lem:fee:constprod:additivity}). 
We leverage this generalised additivity result to show that, in the presence of trading fees, a single large swap achieves a strictly greater gain than two smaller swaps (\Cref{thm:fee:swap-gain:additive}) --- whereas the two strategies yield equal gains in fee-less AMMs. 
Our main economic result is a closed-form solution to the arbitrage problem for fee-adjusted AMMs, namely the problem of determining the swap that maximizes a trader’s gain by exploiting discrepancies between the AMM-implied and external exchange rates.
More specifically, we start by determining a swap value that brings the AMM to an equilibrium state where its internal exchange rate is aligned with the exchange rate given by an external price oracle (\Cref{thm:arbitrage:balance}), and establish its uniqueness (\Cref{lem:arbitrage:balance-unique}).
We then show that this value yields a suboptimal gain, determining an interval of swap values that guarantee a strictly higher gain to the sender (\Cref{thm:fee:equil-vs-gain}).
Our main technical result is a closed-formula solution to the arbitrage problem, namely a swap transaction that obtains the maximum gain (\Cref{thm:fee:max-gain}).
We conclude by showing that this solution is unique (\Cref{lem:arbitrage:optimal-unique}).

\item \textbf{Lean 4 Formalization}: We formalize and prove in Lean 4~\cite{deMoura2021lean4} all the results presented in this work, extending the library presented in \cite{Pusceddu24fmbc}. 
    Our formalization is available online at \url{https://mamboleano.github.io/lean4-amm-fees}, and it consists of ${\sim}$3500 lines of Lean code.
    We additionally provide detailed pen-and-paper proofs, that help outlining the proof strategies implemented in Lean~\cite{Dessalvi25thesis}. 

\end{enumerate}

By providing a machine-checked analysis of the role of trading fees in AMMs, this work contributes to the ongoing effort to improve the robustness and security of DeFi systems.
Because of space constraints, we include several ancillary results in~\cite{Dessalvi25thesis}, and we relegate the discussion of the differences between our model and Uniswap v2 to~\Cref{sec:uniswap-compare}.
\section{AMM model}
\label{sec:amm-model}

In this~\namecref{sec:amm-model}, we refine the operational model of AMMs in~\cite{BCL22lmcs} to introduce the trading fees in the swap rate function, which brings this model closer to the Uniswap v2 protocol~\cite{uniswapv2}.

We assume a set of \emph{token types} $\tokT,\tokTi,\ldots$ representing crypto-assets, and a set of \emph{accounts} $\pmvA,\pmvB,\ldots$, representing the users involved in token exchanges.
An AMM consists of two token reserves, written $\amm{\ammR[0]:\tokT[0]}{\ammR[1]:\tokT[1]}$, where the amounts $\ammR[0],\ammR[1]$ are non-negative reals.
Users interact with AMMs by sending $\ammSwapOp$ transactions. Intuitively, a $\ammSwapOp$ transaction $\actAmmSwapExact{\pmvA}{}{x}{\tokT[0]}{\tokT[1]}$ represents the atomic action by which the user $\pmvA$ pays an amount $x$ of $\tokT[0]$ to the AMM and receives a quantity $y = x \cdot \SX{x,\ammR[0],\ammR[1]}$ of $\tokT[1]$ in return. A central component of the model is therefore the swap rate function $\SX{}$. 
Uniswap v2~\cite{uniswapimpl} and other mainstream AMM implementations~\cite{mooniswapimpl, sushiswapimpl} use the
\keyterm{constant-product swap rate} function: 
\begin{equation}
\label{defi:const-prod}
\SX{x,\ammR[0],\ammR[1]}
= 
\frac{\fee \, \ammR[1]}{\ammR[0] + \fee \, x}
\qquad\text{where } 
\fee \in (0,1]
\end{equation}
where $\ammR[0]$ and $\ammR[1]$ are, respectively, the reserves of the tokens $\tokT[0]$ and $\tokT[1]$ in the current AMM state, and the parameter $\fee$ is the \keyterm{trading fee}. 
Despite the name ``constant-product'', swaps preserve the product $\ammR[0] \cdot \ammR[1]$ of the AMM reserves only in the fee-less case, \ie when $\fee=1$. 
Instead, when $\fee<1$, the product of the reserves strictly increases upon each swap. 
More specifically, if an AMM $\amm{\ammR[0]:\tokT[0]}{\ammR[1]:\tokT[1]}$ evolves into
$\amm{\ammR[0]+x:\tokT[0]}{\ammR[1]-y:\tokT[1]}$ upon a swap, then 
$(\ammR[0]+x) (\ammR[1]-y) > \ammR[0] \ammR[1]$.
Therefore, the amount $y$ of tokens $\tokT[1]$ sent to the user performing the swap is reduced compared to the fee-less case.%
\footnote{These conditions are also specified in Uniswap v2's implementation~\cite{uniswapimpl}, see also \Cref{lst:swap} at page~\pageref{lst:swap}.}
%
The intuition is that when the transaction \mbox{$\actAmmSwapExact{\pmvA}{}{x}{\tokT[0]}{\tokT[1]}$} is fired, the actual number of tokens that are used to re-balance the AMM pair is not $x$, but $\fee \cdot x$ instead. 
This means that $\pmvA$ receives fewer tokens and some units of $\tokT[1]$ token are ``kept'' by the AMM in order to benefit its liquidity providers~(LPs). 
The incentives to LPs are briefly discussed at the end of this section.

\begin{example}
\label{ex:const-prod}
Consider a system $\confG$ containing an AMM $\amm{40 : \tokT[0]}{60 : \tokT[1]}$ and user $\pmvA$'s wallet holding $30:\tokT[0]$ and $20:\tokT[1]$.
We write such state as $\confG = \walA{30 : \tokT[0], \; 20 : \tokT[1]} \mid \amm{40 : \tokT[0]}{60 : \tokT[1]}$. 
Suppose that $\pmvA$ wants to swap $10:\tokT[0]$, and that the trading fee 
in~\eqref{defi:const-prod} is set to $\fee = 0.997$. 
Then, $\pmvA$ expects to receive $y:\tokT[1]$, where: 
\[
    y = 10 \cdot \frac{0.997 \cdot 60}{40 + 0.997 \cdot 10} \approx 11.97
\]
Hence, firing the transaction $\txT = \actAmmSwapExact{\pmvA}{}{10}{\tokT[0]}{\tokT[1]}$ triggers a state transition:
\[
\confG 
\; \xrightarrow{\txT} \;
\confGi = \walA{20 : \tokT[0], \; 31.97 : \tokT[1]} \mid \amm{50 : \tokT[0]}{48.03 : \tokT[1]}
\]
By contrast, by firing the same transaction in an AMM with no trading fee, $\pmvA$ would obtain:
\[
y = 10 \cdot \frac{1 \cdot 60}{40 + 1 \cdot 10} = 
10 \cdot \frac{60}{40 + 10} = 12
\]
and the resulting state would be $\confGii = \walA{20 : \tokT[0], \; 32 : \tokT[1]} \mid \amm{50 : \tokT[0]}{48 : \tokT[1]}$. 
As we can see, in both $\confGi$ and $\confGii$ the AMM has 50 units of $\tokT[0]$ tokens. However, in $\confGi$, the AMM retains a small portion of the $\tokT[1]$ tokens compared to $\confGii$, which results in a slightly worse outcome for the trader $\pmvA$ and a better one for the LPs.  
\hfill\qedex
\end{example}

\paragraph*{Token prices and exchange rates}

We assume a price oracle that determines the price of each token type. 
Technically, we model this oracle as a function $\exchO{} : \TokU[0] \rightarrow \Rpos$, where $\TokU[0]$ is the universe of token types.

Based on the prices provided by this price oracle, we define the \keyterm{external exchange rate} $\X{\tokT[0], \tokT[1]}$ between two token types $\tokT[0]$ and $\tokT[1]$ as the ratio between their prices:
\[
    \X{\tokT[0], \tokT[1]} = \frac{\exchO{}(\tokT[0])}{\exchO{}(\tokT[1])}
\]
This represents the number of units of $\tokT[1]$ tokens a user can buy with 1 unit of $\tokT[0]$ using \eg a centralized exchange platform. 
Note that the external exchange rate only depends on the prices given by the oracle, neglecting the state of AMMs.

The \keyterm{internal exchange rate} $\X[\confG]{\tokT[0], \tokT[1]}$, by contrast, is determined by the swap rate function $\SX{}$ and by the reserves of an AMM pair in $\confG$ handling the token pair $\setenum{\tokT[0],\tokT[1]}$. 
We assume that each state $\confG$ can contain at most one AMM for each token pair.
Its interpretation mirrors that of the external exchange rate: for a \emph{very small} amount of input tokens $x$, a user swapping $x$ units of $\tokT[0]$ expects to receive $x \cdot \X[\confG]{\tokT[0], \tokT[1]}$ units of $\tokT[1]$ tokens. 
More formally:
\[
    \X[\confG]{\tokT[0], \tokT[1]} = \lim_{x \rightarrow 0} \SX{x, \ammR[0], \ammR[1]} \qquad 
    \text{if } \confG \text{ contains } \amm{\ammR[0]:\tokT[0]}{\ammR[1]:\tokT[1]}  
\]


\begin{example}\label{ex:exch-rates}
    Let $\confG = \walA{30 : \tokT[0], \; 20 : \tokT[1]} \mid \amm{40 : \tokT[0]}{60 : \tokT[1]}$. 
    Suppose that the external prices of $\tokT[0]$ and $\tokT[1]$ are $\exchO{\tokT[0]} = 4$ and $\exchO{\tokT[1]} = 5$. 
    Hence, the external exchange rate is: 
    \[
        \X{\tokT[0],\tokT[1]} = \frac{4}{5} = 0.8
    \]
    Assuming that we are using the constant-product swap rate with trading fee $\fee = 0.997$, the internal exchange rate computed in the state $\confG$ is: 
    \begin{align*}
    \X[\confG]{\tokT[0], \tokT[1]} 
    & = \lim_{x \rightarrow 0} \SX{x, \ammR[0], \ammR[1]}
    \; = \; 
    \lim_{x \rightarrow 0} \SX{x, 40, 60}
    \; = \;
    \lim_{x \rightarrow 0} \frac{0.997 \cdot 60}{40 + 0.997 \cdot x}
    \\[5pt] 
    & =
    \frac{0.997 \cdot 60}{40}
    \; = \; 1.4955
    \end{align*}
    Therefore, in this state the external and internal exchange rates are not aligned.
    We will show later that rational users are incentivized to execute swaps that align internal and external exchange rates, a practice called \keyterm{arbitrage}.
    \hfill\qedex
\end{example}

\paragraph*{Users' Gain}

The \keyterm{gain} of a user $\pmvA$ from executing a transaction $\txT$ in a state $\confG$, denoted by $\gain[\confG]{\pmvA}{\txT}$, is defined as the difference between the wealth of $\pmvA$ after and before the transaction is executed.
Wealth is measured as the weighted sum of the token balances in $\pmvA$'s wallet, where the weights are given by the external token prices.

\begin{example}
\label{ex:const-prod:gain}
Recall the scenario from~\Cref{ex:const-prod}.
The wealth of $\pmvA$ in $\confG$ is $30 \cdot 4 + 20 \cdot 5 = 220$, while that in $\confGi$ is $20 \cdot 4 + 31.97 \cdot 5 = 239.85$.
Therefore, $\pmvA$'s gain upon performing the transaction
\mbox{$\txT = \actAmmSwapExact{\pmvA}{}{10}{\tokT[0]}{\tokT[1]}$} in $\confG$ is:
\[
    \gain[\confG]{\pmvA}{\txT}
    = 239.85 - 220 
    = 19.85
\]
Note that $\pmvA$ had a profit from the swap,  
and the internal exchange rate shifted from $\X[\confG]{\tokT[0], \tokT[1]} = 1.4955$ to $\X[\confGi]{\tokT[0], \tokT[1]} \approx 0.958$, making it closer to the external rate $\X{\tokT[0], \tokT[1]} = 0.8$.
In general, we expect that swaps that reduce the gap between internal and external exchange rate always yield a profit.
We will formalize this intuition in~\Cref{sec:economic-properties}.
\hfill\qedex
\end{example}

\paragraph*{Liquidity provision}
We implicitly assume the presence of \emph{minted tokens}. For example, the minted token $\tokM{\tokT[0]}{\tokT[1]}$ represents a share of the AMM $\amm{\ammR[0]:\tokT[0]}{\ammR[1]:\tokT[1]}$. Normally, a user would gain these shares by providing liquidity to the AMM; however, here we only model trading users, leaving the analysis of the impact of fees on liquidity providers as possible future work. Hence, in the following statements, we assume that the supply of the minted tokens is always positive (\ie, that at some point liquidity has been provided to the AMM), and that no trader holds all of the minted token supply. These assumptions rule out unrealistic scenarios in which, for example, the trader has complete control over the AMM shares.

\section{Economic properties of AMMs with trading fees}
\label{sec:economic-properties}

In this~\namecref{sec:economic-properties}, we study the economic properties of AMMs with trading fees. Our goal is to analyze how trading fees play a role on the incentives of users and the behavior and properties of swap functions, with keen focus on the constant-product swap rate function~\eqref{defi:const-prod}, which is the function used in several real-world AMMs implementations~\cite{mooniswapimpl,sushiswapimpl,uniswapimpl}.




\paragraph*{Properties of the costant-product swap rate function}
\label{sec:properties-const}

AMMs using the constant-product swap rate function (hereaafter, referred to as \emph{CPMMs}) are \emph{output-bounded}~\cite{BCL22lmcs}. 
Namely, they always have enough output tokens $\tokT[1]$ to send to the user who performs a  
$\actAmmSwapExact{}{}{x}{\tokT[0]}{\tokT[1]}$, for any value of $x$.
In particular, output-boundedness guarantees that the AMM will always hold a strictly positive amount of each token type.

\citeLean{AMMLib/FeeVersion/Constprod.html\#SX.fee.constprod.outputbound} 
\begin{lemma}
\label{lem:fee:constprod:output-bound}
CPMMs are \emph{output-bounded}, \ie
for all $x \geq 0$ and $\ammR[0], \ammR[1] > 0$:
\[
x \cdot \SX{x,\ammR[0],\ammR[1]} < \ammR[1]
\]
\end{lemma}

CPMMs are also \emph{strictly monotonic}~\cite{BCL22lmcs}.
Namely, for a swap action $\actAmmSwapExact{}{}{x}{\tokT[0]}{\tokT[1]}$,
the swap rate (strictly) increases if we (strictly) decrease the input amount $x$ or the reserves of $\tokT[0]$, or if we (strictly) increase the reserves of $\tokT[1]$.
This monotonicity property plays a crucial role in establishing the existence and optimality of arbitrage transactions.

\citeLean{AMMLib/FeeVersion/Constprod.html\#SX.fee.constprod.strictmono} 
\begin{lemma}
\label{lem:fee:constprod:strict-mono}
CPMMs are \emph{strictly monotonic}, \ie
for $i \in \setenum{0,1,2}$ and $\lhd_i \in \setenum{<,\leq}$:
\[
    x' \lhd_0 x,\; \ammRi[0] \lhd_1 \ammR[0],\; \ammR[1] \lhd_2 \ammRi[1]
    \implies
    \SX{x,\ammR[0],\ammR[1]}
    \lhd_3
    \SX{x',\ammRi[0],\ammRi[1]}
\]
where: \
$\lhd_3 = \leq$ if $\lhd_i = \,\leq$ for $i \in \setenum{0,1,2}$,
and $\lhd_3 = <$ otherwise.
\end{lemma}

The following property is useful to measure the effect of splitting a single swap in two swaps of smaller amounts. In particular, it determines the factor $\Z x y {r_0} {r_1}$ between the swap rate of a single transaction swapping $x+y$ tokens and the rates $\alpha$ and $\beta$ of two transactions (swapping first $x$, then $y$) in a CPMM with reserves $r_0,r_1$.


\citeLean{AMMLib/FeeVersion/Constprod.html\#SX.fee.constprod.extended_additivity} 
\begin{lemma}[Additivity of swap rate]
\label{lem:fee:constprod:additivity}
CPMMs are \emph{additive}, \ie:
\[
    \SX{x+y,\ammR[0],\ammR[1]} = 
    \frac{\valSXa x + \valSXb y}{x+y} \cdot \Z x y {r_0} {r_1}
\]
where 
$\valSXa = \SX{x,\ammR[0],\ammR[1]}$, 
$\valSXb = \SX{y,\ammR[0]+x,\ammR[1]-\valSXa x}$, and:
\begin{align*}
    \Z x y {r_0} {r_1} = 
    \frac
    {\Big ( (\fee \ammR[1] x)(\ammR[0] + \fee x + \fee y) + (\fee \ammR[1] \ammR[0] y)\Big ) \cdot (\ammR[0] + x + \fee y)}
    {(\ammR[0] + \fee x + \fee y) \cdot \Big ( (\fee \ammR[1] x)(\ammR[0] + x + \fee y) + (\fee \ammR[1] \ammR[0] y)\Big )}
\end{align*}
In particular, if $\fee=1$ then $\Z x y {r_0} {r_1} = 1$. 
\end{lemma}

In CPMMs without trading fees, the gain obtained by swapping $x+y$ tokens is equal to the sum of the gains obtained from two consecutive swaps of $x$ and $y$ tokens, respectively~\cite{BCL22lmcs}. 
In the presence of a trading fee, however, this additivity property no longer holds: technically, this follows by the fact that the factor $\Z x y {r_0} {r_1}$ is strictly positive when $\fee < 1$. 
Intuitively, this implies that a user achieves a strictly higher gain by executing a single large swap rather than splitting the transaction into multiple smaller swaps.


\begin{example}
\label{ex:fee:constprod:additivity}
    Let $\confG = \walA{300 : \tokT[0], 200 : \tokT[1]} \mid \amm{400 : \tokT[0]}{600 : \tokT[1]}$. Now suppose that the prices of $\tokT[0]$ and $\tokT[1]$ are $\exchO{\tokT[0]} = 4$ and $\exchO{\tokT[1]} = 5$. Let \mbox{$\txT[0] = \actAmmSwapExact{\pmvA}{}{100}{\tokT[0]}{\tokT[1]}$} and let 
    $\fee = 0.997$. 
    By executing $\txT[0]$ in $\confG$, $\pmvA$ would obtain 
    $y = 100 \cdot \SX{100, 400, 600} = 119.712:\tokT[1]$, and so: 
    \[
        \confG
        \xrightarrow{\txT[0]}
        \confGi = \walA{200 : \tokT[0], 319.712 : \tokT[1]} \mid \amm{500 : \tokT[0]}{480.288 : \tokT[1]}
    \]
    Now, suppose that $\pmvA$ --- rather than firing $\txT[0]$ --- splits this transaction into two transactions, 
    \mbox{$\txT[1] = \actAmmSwapExact{\pmvA}{}{40}{\tokT[0]}{\tokT[1]}$} and \mbox{$\txT[2] = \actAmmSwapExact{\pmvA}{}{60}{\tokT[0]}{\tokT[1]}$}. 
    We would have that:
    \begin{align*}
    \confG 
    \xrightarrow{\txT[1]}
    & \; \confG[1] = \walA{260 : \tokT[0], 254.397 : \tokT[1]} \mid \amm{460 : \tokT[0]}{545.603 : \tokT[1]}
    \\
    \xrightarrow{\txT[2]} 
    & \; \confG[2] = \walA{200 : \tokT[0], 319.696 : \tokT[1]} \mid \amm{500 : \tokT[0]}{480.304 : \tokT[1]}
    \end{align*}
    We see that the split did not benefit $\pmvA$, as she received less tokens than what she would have received by firing the transaction $\txT[0]$ directly instead of splitting it in $\txT[1]$ and $\txT[2]$. 
    \hfill\qedex
\end{example}

The impact on a user's gain of splitting a large swap into two smaller swaps is characterized by the following theorem:

\citeLean{AMMLib/FeeVersion/Additivity.html\#SX.fee.additive_gain}
\begin{theorem}[Additivity of swap gain]
\label{thm:fee:swap-gain:additive}
  Consider a CPMM $\amm{\ammR[0]:\tokT[0]}{\ammR[1]:\tokT[1]}$ in state $\confG$ and let $\txT(x) = \actAmmSwapExact{\pmvA}{}{x}{\tokT[0]}{\tokT[1]}$. Let $\confGi$ and $\confGii$ be such that $\confG \xrightarrow{\txT(x_0)} \confGi \xrightarrow{\txT(x_1)} \confGii$.
  \smallskip
  If $\fee < 1$ then there exists $\epsilon_{\fee} > 0$ such that:
  \[
    \gain[\confG]{\pmvA}{\txT(x_0+x_1)} 
    \; = \;
    \gain[\confG]{\pmvA}{\txT(x_0)} + \gain[\confGi]{\pmvA}{\txT(x_1)} + \epsilon_{\fee}
  \]
\end{theorem}

The term $\epsilon_{\fee}$ captures the correction induced by the presence of trading fees. Since its definition is algebraically heavy and does not provide any additional insight, we refer to~\cite{Dessalvi25thesis} and to our Lean library for its formalization. 
What is relevant, though, is that in the presence of a fee ($\fee < 1$), the factor $\epsilon_{\fee}$ is strictly positive, reflecting the fact that splitting a swap into multiple transactions results in a decreased gain for the trader. On the other hand, if $\fee = 1$, then also $\epsilon_{\fee}=0$. Therefore, our result is coherent with ~\cite{BCL22lmcs} in the fee-less case. 



\paragraph*{Arbitrage}

We now address the problem of finding the optimal amount of tokens that a trader should swap to maximize their gain. 
In CPMMs without trading fees, the optimal strategy consists in executing a swap that aligns the internal exchange rate to the external one~\cite{BCL22lmcs}, thereby bringing the AMM to a so-called \emph{equilibrium} state. 
Trying to follow the same strategy in the presence of trading fees, the first step is determining the swap value that brings the AMM to the equilibrium.
This is given by the following~\namecref{thm:arbitrage:balance}:

\vbox{
\citeLean{AMMLib/FeeVersion/Arbitrage.html\#SX.fee.arbitrage.constprod.equil_value} 
\begin{theorem}[Equilibrium value]
\label{thm:arbitrage:balance}
  Consider a CPMM $\amm{\ammR[0]:\tokT[0]}{\ammR[1]:\tokT[1]}$ in state $\confG$, and let:
  \begin{equation}
    \label{eq:arbitrage:balance}
    x_0 
    \; = \;
    \frac
        {-\sqrt{\exchO{\tokT[0]}} \ammR[0] (1 + \fee) + \sqrt{\ammR[0]} \sqrt{\exchO{\tokT[0]} \ammR[0] (-1 + \fee)^2 + 4 \exchO{\tokT[1]} \ammR[1] \fee^2}}
        {2 \sqrt{\exchO{\tokT[0]}} \fee}
  \end{equation}
  If $\txT(x_0) = \actAmmSwapExact{\pmvA}{}{x_0}{\tokT[0]}{\tokT[1]}$ is enabled in $\confG$ and $\fee < 1$, then
  \(
      \X{\tokT[0], \tokT[1]} = \X[\confGi]{\tokT[0], \tokT[1]} 
  \), 
  where $\confGi$ is the state reached from $\confG$ upon executing $\txT(x_0)$.
\end{theorem}
} 

\begin{figure}[t]
  \centering
  \begin{minipage}[t]{0.45\textwidth}
    \centering
    \includegraphics[width=\linewidth]{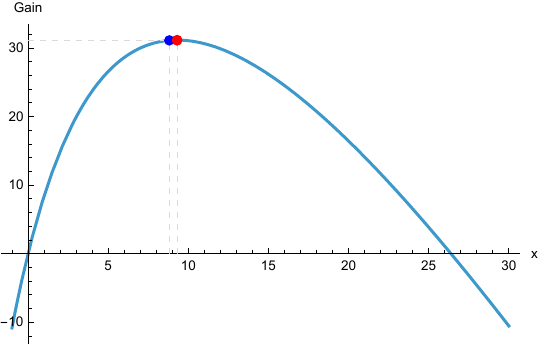}
    \caption{$\gain[{\confG}]{\pmvA}{\txT(x)}$}
    \label{fig:gain-xEquil}
  \end{minipage}\hfill%
  \begin{minipage}[t]{0.45\textwidth}
    \centering
    \includegraphics[width=\linewidth]{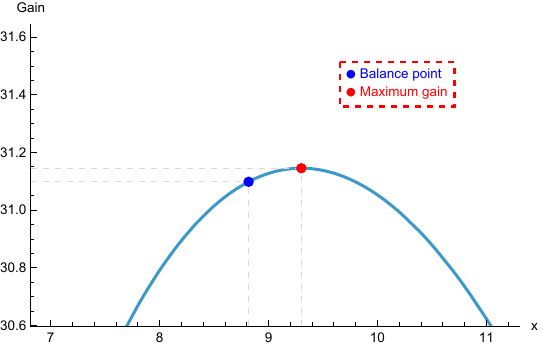}
    \caption{$\gain[{\confG}]{\pmvA}{\txT(x)}$ zoomed.}
    \label{fig:gain-zoomed}
  \end{minipage}
\end{figure}

The following lemma states that, when the value $x_0$ given by~\Cref{thm:arbitrage:balance} satisfies the enabledness condition, it is the only one that brings the AMM to the equilibrium.

\citeLean{AMMLib/FeeVersion/Arbitrage.html\#SX.fee.arbitrage.constprod.solution_equil_unique} 
\begin{lemma}[Uniqueness of equilibrium swap value]
\label{lem:arbitrage:balance-unique}
  Consider a CPMM $\amm{\ammR[0]:\tokT[0]}{\ammR[1]:\tokT[1]}$ in state $\confG$, and let $\txT(x) = \actAmmSwapExact{\pmvA}{}{x}{\tokT[0]}{\tokT[1]}$.
  If $\txT(x_0)$ and $\txT(x_1)$ bring the AMM to the equilibrium state from $\confG$,
  then $x_0 = x_1$.
\end{lemma}

We now address the problem of determining if the swap value that brings an AMM to the equilibrium also maximizes the user's gain.
While this equivalence holds for CPMMs without trading fees~\cite{BCL22lmcs}, the following example shows that it fails when $\fee < 1$. 

\begin{example}
    \label{ex:gain-equil}
    Consider a CPMM in the state
    $\confG = \walA{30 : \tokT[0], 20 : \tokT[1]} \mid \amm{10 : \tokT[0]}{30 : \tokT[1]}$. 
    Assume a trading fee $\fee = 0.9$, and external token prices $\exchO{\tokT[0]} = 4$ and $\exchO{\tokT[1]} = 5$. 
    Let \mbox{$\txT(x) = \actAmmSwapExact{\pmvA}{}{x}{\tokT[0]}{\tokT[1]}$}, and let $x_0$ be the swap amount determined by~\eqref{eq:arbitrage:balance}, \ie the value that brings the AMM to equilibrium. 
    The gain obtained by $\pmvA$ from executing $\txT(x_0)$ in $\confG$ is: 
    \[  \numberthis \label{eq:ex:gain-equil}
        \gain[{\confG}]{\pmvA}{\txT(x_0)} \approx 31.098
    \]
    In the fee-less setting of~\cite{BCL22lmcs}, this value would coincide with the maximum achievable gain from any swap executed in state $\confG$. 
    In contrast, this is no longer the case in the presence of trading fees. \Cref{fig:gain-xEquil} plots the gain function $\gain[{\confG}]{\pmvA}{\txT(x)}$. 
    The function is concave, \ie it has a critical point which is also the absolute maximum for the function. 
    However, a closer inspection (see~\Cref{fig:gain-zoomed}) reveals that the gain at $x_0$ (in blue) is not exactly at the top of the curve. 
    Indeed, evaluating the gain at a slightly larger swap amount $x_1 = x_0 + 0.3$ yields: 
    \[
        \gain[{\confG}]{\pmvA}{\txT(x_1)} \approx 31.138
    \]
    which is strictly greater than the gain in~\eqref{eq:ex:gain-equil}.
    This example demonstrates that, when $\fee < 1$, the equilibrium-inducing swap does not necessarily maximize the user's gain.
    \hfill\qedex
\end{example}

Although the equilibrium-inducing swap is suboptimal, it remains very close to the optimum. It is therefore still meaningful to compare the gain achieved by this swap amount with that obtained from any other swap value, as formalized by the~\namecref{thm:fee:equil-vs-gain} below.

\citeLean{AMMLib/FeeVersion/Arbitrage.html\#SX.fee.arbitrage.constprod.equil_value_solution_arbitrage} 
\begin{theorem}[Equilibrium value vs gain]
\label{thm:fee:equil-vs-gain}  
  Let $\confG = \walA{\tokBal[\pmvA]} \mid \amm{\ammR[0]:\tokT[0]}{\ammR[1]:\tokT[1]} \mid \confD$ 
  be such that $\stdSupply > 0$ and $\stdWallMint < \stdSupply$
  and let $\txT(x) = \actAmmSwapExact{\pmvA}{}{x}{\tokT[0]}{\tokT[1]}$.
  Let $x_0$ be the value determined in~\eqref{eq:arbitrage:balance}.
  If $\fee < 1$, then, whenever $\txT(x_0)$ and $\txT(x)$ are enabled in $\confG$:
  \begin{gather*}
    \forall x < x_0 
    \; : \;
    \gain[{\confG}]{\pmvA}{\txT(x_0)}
    \; > \;
    \gain[{\confG}]{\pmvA}{\txT(x)}
    \\
    \forall x > x_0 
    \; : \;
    \Big( 
    \gain[{\confG}]{\pmvA}{\txT(x_0)}
    \; > \;
    \gain[{\confG}]{\pmvA}{\txT(x)} \iff x > \frac{x_0}{\fee}
    \Big)
   \end{gather*}
\end{theorem}

This behaviour is consistent with~\Cref{ex:gain-equil} and the plots in \Cref{fig:gain-xEquil,fig:gain-zoomed}.
Specifically, there exists a small interval immediately to the right of $x_0$ where $\pmvA$'s gain increases relative to that obtained by swapping $x_0$ tokens. 
This is the interval $(x_0,\nicefrac{x_0}{\fee}]$. 
Consequently, the remaining task is to identify the swap amount within this interval which maximizes the gain.

\citeLean{AMMLib/FeeVersion/Maximal.html\#SX.fee.constprod.x_max_gain} 

\begin{theorem}[Max Gain Value]
\label{thm:fee:max-gain}
  Consider a CPMM $\amm{\ammR[0]:\tokT[0]}{\ammR[1]:\tokT[1]}$
  in a state $\confG$ where $\pmvA$ does not hold all minted tokens $\setenum{\tokT[0],\tokT[1]}$.
  Let $\txT(x) = \actAmmSwapExact{\pmvA}{}{x}{\tokT[0]}{\tokT[1]}$.
  Let $x_0$ be a swap value bringing the AMM to equilibrium, 
  and let:
  \begin{align*}
    & x_{max} = x_0 + \frac
        {-\sqrt{\exchO{\tokT[0]}} \ammR[0] - \sqrt{\exchO{\tokT[0]}} \fee x_0 + \sqrt{\exchO{\tokT[1]} \fee \ammR[0] \ammR[1]}}
        {\sqrt{\exchO{\tokT[0]}} \fee}
    &&    
    \begin{array}{l}
    \text{where }
    \confG \xrightarrow{\txT(x_0)} \confGi
    \text{ and }
    \\
    \X[\confGi]{\tokT[0],\tokT[1]} = \X{\tokT[0],\tokT[1]}
    \end{array}
  \end{align*}
  If $\txT(x_{\it max})$ is enabled in $\confG$ and $\fee < 1$, then
  $\txT(x_{\it max})$ maximizes $\pmvA$'s gain:
  \[
    \forall x \neq x_{\it max}
    \; : \;
    \gain[{\confG}]{\pmvA}{\txT(x_{\it max})}
    \; > \;
    \gain[{\confG}]{\pmvA}{\txT(x)}
  \]
\end{theorem}

Finally, it is natural to ask if $x_{\it max}$ is the unique value that maximizes the trader's gain. 
This is established by the following lemma, proved by contradiction. 

\citeLean{AMMLib/FeeVersion/Maximal.html\#SX.fee.constprod.x_max_unique} 
\begin{lemma}[Optimal swap value uniqueness]
\label{lem:arbitrage:optimal-unique}
    Consider a CPMM $\amm{\ammR[0]:\tokT[0]}{\ammR[1]:\tokT[1]}$ in state $\confG$ and let $\txT(x) = \actAmmSwapExact{\pmvA}{}{x}{\tokT[0]}{\tokT[1]}$. If $\fee < 1$ and $\txT(x_{max})$ and $\txT(x_{max'})$ both maximize the gain of $\pmvA$ from $\confG$, then $x_{max} = x_{max'}$.
\end{lemma}

\section{Related Work}
\label{sec:related}

The literature on Automated Market Makers is extensive, as AMMs are one of the core building blocks of Decentralized Finance~\cite{Kitzler23tweb,Werner22aft,Xu21sok}. 
In this section, we focus primarily on comparing our work with the foundational studies on which it builds, as well as with other works that explicitly address fee mechanisms in AMMs. 

The work \cite{BCL22lmcs} introduces an abstract operational model of AMMs that forms the foundation of our work. 
While most our formalization and results focus on the constant-product swap rate function, the analysis in~\cite{BCL22lmcs} deliberately abstracts from any specific instantiation of the swap rate function, instead identifying conditions that guarantee ideal structural properties of AMMs, such as additivity and reversibility of swaps. 
The main difference between our work and~\cite{BCL22lmcs} lies in the treatment of trading fees, which are not modelled or analysed in \cite{BCL22lmcs}. 
While the structural properties in \cite{BCL22lmcs} are preserved in our model in the fee-less case $\fee=1$, when $\fee<1$ some key properties need to be adapted.
In particular: 
\begin{enumerate}
    \item \textbf{Additivity}: Consider a user executing a swap of $x:\tokT[0]$ followed by another swap of $y:\tokT[0]$ on the same AMM.
    The exchange rates corresponding to these swaps are given by:  
    \[
        \valSXa = \SX{x, \ammR[0], \ammR[1]} \text{ and } \valSXb = \SX{y, \ammR[0] + x, \ammR[1] - \valSXa x}
    \]
    where $\ammR[0]$ and $\ammR[1]$ are the reserves of the AMM pair when the first swap is performed. 
    In the framework of~\cite{BCL22lmcs}, a swap rate function $\SX{}$ is defined to be additive if: 
    \[
    \SX{x + y, \ammR[0], \ammR[1]} = \frac{\valSXa x + \valSXb y}{x + y}
    \]
    When this equation holds, swapping an amount of tokens $(x + y):\tokT[0]$ is equivalent to swapping first $x:\tokT[0]$ and subsequently $y:\tokT[0]$.
    Indeed, the output amount of the single swap with input $(x+y):\tokT[0]$ is:
    \[
    (x+y) \cdot \SX{x + y, \ammR[0], \ammR[1]} 
    = 
    (x+y) \cdot \frac{\valSXa x + \valSXb y}{x + y}
    = 
    \valSXa x + \valSXb y
    \]
    which is the sum of the output amounts of two subsequent swaps of $x:\tokT[0]$ and $y:\tokT[0]$.
    We have show in~\Cref{ex:fee:constprod:additivity} that the fee-adjusted constant-product swap rate does not respect this additivity property.
    Then, in~\Cref{lem:fee:constprod:additivity} we devised a generalized additivity notion that takes into account the trading fee, by multiplying the previous formula by the factor $\Z x y {\ammR[0]} {\ammR[1]}$.
    Building upon this lemma, \Cref{thm:fee:swap-gain:additive} generalises Lemma 5.7 in~\cite{BCL22lmcs}, stating that, the presence of trading fees, performing two subsequent swaps on the same AMM pair yields a \emph{smaller} gain than performing a single large swap.

    \item \textbf{Reversibility}: Let $\valSXa = \SX{x, \ammR[0], \ammR[1]}$, where $\ammR[0]$ and $\ammR[1]$ are the current reserves of an AMM pair. 
    The work \cite{BCL22lmcs} defines a swap rate function to be reversible if, for all $x>0$: 
    \[
        \SX{\valSXa x, \ammR[1] - \valSXa x, \ammR[0] + x} = \frac{1}{\valSXa}
    \]
    When this condition holds, a swap of $x:\tokT[0]$ for $y:\tokT[1]$ immediately followed by a swap of all the received tokens $\tokT[1]$ in the opposite direction, restores the AMM to the original state.
    This reversibility property holds in the fee-less setting (Theorem 5.9 in~\cite{BCL22lmcs}), but it is no longer valid in the presence of a trading fee $\fee < 1$.
    Since trading fees make the product of the AMM reserves strictly increase after each swap, a trader attempting to reverse the effect of a swap always  incur in a net loss.
    The absence of reversibility has significant methodological implications. Several results in~\cite{BCL22lmcs} rely critically on this property, including Theorem~6.3, which establishes the optimality of the arbitrage transaction.
    In contrast, our corresponding results (\Cref{thm:fee:max-gain,lem:arbitrage:optimal-unique}) are proved using different proof strategies that do not rely on reversibility of the swap rate function, and are therefore applicable also in the presence of trading fees.

    \item \textbf{Equilibrium \emph{vs.} Arbitrage}:
    Recall that an AMM is in an equilibrium state when its internal exchange rate equals the external rate given by a price oracle.
    In the fee-less setting, the swap amount that brings an AMM to equilibrium coincides with the arbitrage solution, \ie, it is also the value that maximizes the trader's profit (Theorem~6.3 in~\cite{BCL22lmcs}). 
    This equivalence no longer holds in the presence of a trading fee. 
    We have shown that the equilibrium-inducing swap amount, determined in~\Cref{thm:arbitrage:balance}, differs from the profit-maximising amount, which is instead characterised in~\Cref{thm:fee:max-gain}.
    To the best of our knowledge, ours is the first work that, in a fee-adjusted AMM model, formally compares the swap amount that aligns internal to external prices  versus the swap amount that maximizes the individual arbitrage profit.
    
\end{enumerate}

Our Lean 4 implementation builds on the AMM library presented in~\cite{Pusceddu24fmbc}, which formalizes the core functionalities of AMMs 
and provides machine-checked proofs of key economic properties of CPMMs studied in~\cite{BCL22lmcs}, including the arbitrage solution in the fee-less setting.
We extend this library with trading fees and provide machine-checked proofs of  the corresponding fee-adjusted economic properties.
In contrast, the formalization of~\cite{Pusceddu24fmbc} also models liquidity provision and withdrawal, enabling the analysis of liquidity providers' incentives.

The work \cite{Nielsen23cpp} proposes a framework for developing and formally verifying AMMs in the Coq proof assistant. In this approach, an AMM is modelled as a composition of smart contracts, formalized as Coq functions on top of ConCert~\cite{Annenkov20cpp}, a mechanized framework for blockchain and smart contract verification.
Although both~\cite{Nielsen23cpp} and our work formalise AMM within a proof assistant, their objectives differ substantially. 
The formalization in~\cite{Nielsen23cpp} closely follows a concrete AMM implementation (Dexter2) and yields a deployable contract that is provably consistent with its Coq specification. 
By contrast, we adopt a more abstract model, aimed at identifying and formally proving properties that must hold for any implementation conforming to the specification.
As a result, while both works model trading fees, only ours analyses their impact on key economic properties such as arbitrage and equilibrium.

A different formalization of AMMs is proposed in~\cite{BMZ25arxiv}, within a general Lean 4 library aimed at the formalization of Maximal Extractable Value (MEV).
This is a class of attacks where adversaries with transaction sequencing privileges can ``extract'' value from honest users' transactions by strategically reordering them in the blockchain, potentially front-running or back-running the victim's transactions with adversarial ones~\cite{Babel23clockwork}.
AMMs are notoriously prone to MEV attacks. In particular, the so-called \emph{sandwich attack} allows the adversary to extract value from virtually any swap transaction~\cite{Zhou21high,BCL22fc}.
The Lean formalization in~\cite{BMZ25arxiv} provides a machine-checked proof of the optimality of sandwich attacks in an AMM model without trading fees.  
Extending this result to account for trading fees would likely require a substantial amount of additional work, since --- even in the fee-less setting --- the space of attack strategies that an adversary must consider is surprisingly large.  

The work~\cite{Milionis2024fc} studies ``constant-function'' AMMs Makers~\cite{Angeris20aft}, whereas we focus on ``constant-product'' AMMs, a subclass of this broader family. The model in~\cite{Milionis2024fc} differs substantially from ours: building on~\cite{Milionis2022LVR}, trading fees are analyzed in a stochastic setting with discrete Poisson block arrivals, leading to results that emphasize the long-term impact of fees on arbitrageurs' behavior. By contrast, our analysis focusses on single-swap incentives and equilibrium properties. Finally, unlike our work, the results in~\cite{Milionis2024fc} are established via pen-and-paper proofs and are not supported by a machine-checked formalization.

Several works address the role of fees in AMMs from perspectives complementary to ours. In particular, \cite{Fritsch2022EconomicsAMM} investigates the so-called \emph{protocol fee}, namely the portion of the trading fee that is diverted to the protocol and used to generate revenue for governance token holders. That work also analyzes the incentives of both liquidity providers and traders.
The modeling assumptions in~\cite{Fritsch2022EconomicsAMM} differ significantly from those adopted here: their framework incorporates features such as sticky liquidity and trade volume, whereas our work focuses on a minimal operational model of AMMs amenable to formal, machine-checked reasoning.


In summary, our work extends this line of research by incorporating trading fees into both an operational AMM model and its Lean 4 implementation, thereby bringing the framework closer to real-world deployments such as Uniswap v2 and contributing to the development of more secure and reliable DeFi systems.
\section{Conclusions}

We have presented a rigorous and machine-checked analysis of AMMs with trading fees, built on the abstract operational model presented in \cite{BCL22lmcs} and its Lean 4 formalization in \cite{Pusceddu24fmbc}, here extended with the following main contributions. 
We have started in~\Cref{sec:amm-model} by introducing the trading fee $\fee$ as a parameter of the constant-product swap rate function:
\[
    \SX{x, \ammR[0], \ammR[1]} = \frac{\fee \ammR[1]}{\ammR[0] + \fee \cdot x}
\]
We have then shown that, in the presence of a trading fee $\fee < 1$,
the product of the AMM reserves strictly increases after a swap, \ie
\(
     (\ammR[0] + x)(\ammR[1] - y) > \ammR[0] \ammR[1] 
\).
This implies that the liquidity providers' shares of the AMM also increase compared to the fee-less model.

We have studied the impact of trading fees on key economic properties of AMMs. 
In particular, we show that the fee-adjusted constant-product swap rate remains \emph{output-bounded} and \emph{strictly monotonic}. 
These properties ensure that the AMM reserves cannot be fully drained by a swap action and that the swap rate behaves monotonically \wrt its inputs (\eg, increasing the input amount in a swap results in an increase of the output amount).
We generalise the \emph{additivity} property of~\cite{BCL22lmcs}, showing that  a single large swap yields a strictly greater gain than splitting the same total amount into multiple smaller trades (\Cref{thm:fee:swap-gain:additive}).

Finally, we have investigated the arbitrage problem and how to solve it. We first have determined the swap amount bringing an AMM to equilibrium  (\Cref{thm:arbitrage:balance}). 
We then have shown that in the presence of fees, this value is not the optimal one, as (\Cref{thm:fee:equil-vs-gain}), and we find a closed formula for such optimal value (\Cref{thm:fee:max-gain}). 
Additionally, we have established the uniqueness of both the equilibrium value and the optimal one (\Cref{lem:arbitrage:balance-unique,lem:arbitrage:optimal-unique}).
    
All results presented in this work have been formalised and machine-checked in the Lean~4 proof assistant.
This guarantees the mathematical correctness of our development and extends the existing AMM library, thereby providing a foundation for future work on the formal verification of AMMs.  
For readability, we additionally provide (human-checked) pen-and-paper proofs of our results in~\cite{Dessalvi25thesis}. 


\paragraph*{Challenges of formalizing AMMs in Lean 4}
\label{sec:challenges-lean4}


Although we were able to prove all of the lemmas and theorems above, the current Lean 4 library that implements this model and all the relative proofs has some limitations.

The type used to represent positive amounts in the model, \ie some $x \in \Rpos$, poses significant challenges. This type is defined as a \emph{subtype} of the already existing --- and well supported --- Reals $\mathbb{R}$. 
This approach has advantages, such as guaranteeing that the results of some lemmas are strictly positive (for example, it ensures that for an initialized AMM the reserves are always strictly positive), but it also comes with a few challenges. For example, positive reals by definition do not have a zero element, hence they cannot form a ring, since it lacks additive identity and inverse. 
This might seem a minor issue, but it greatly impacts some of the Lean 4 tactics that are typically used to perform algebraic manipulation (\eg \codelst{ring-nf}). This means that automation in proofs is limited and not always useful.

Another issue that arose during the development of the Lean proofs concerns the use of simplification lemmas. There are a number of already existing simplification lemmas that are not always meant to be used by the Lean 4 simplifier. 
This can negatively affects proof automation, as the prover may get stuck in infinite rewriting loops caused by inappropriate simplification rules. 
Investigating more principled criteria for selecting and organizing simplification lemmas could therefore lead to significant improvements in automation.

Finally, in the proof of~\Cref{thm:fee:max-gain}, we note that a natural proof strategy would be to follow a classical function-analysis argument. 
In particular, since the gain function is concave, one could establish optimality by showing that a local maximum is also a global maximum.
This reasoning is formalized in Mathlib 4 by the theorem \codelst{of_isLocalMaxOn_of_concaveOn}\footnote{\url{https://leanprover-community.github.io/mathlib4_docs/Mathlib/Analysis/Convex/Extrema.html\#IsMaxOn.of\_isLocalMaxOn\_of\_concaveOn}}. 
However, to the best of our knowledge, this approach is currently not directly applicable within 
our Lean 4 formalization.
The problem is that the Mathlib 4 theorem cited before specifically requires the function domain to be an additive commutative monoid, while the gain function in our model is defined in the domain $\Rpos$. 
As a result, the required concavity-based argument cannot be applied directly. While alternative encodings or workarounds may be possible, we preferred to find an alternative proof strategy. 

All the challenges described above can be thought of as trade-offs associated with using Lean 4. On the positive side, Lean 4 allows one to formalize models and properties of such in a precise and uniform way, and to verify that all stated properties hold without gaps. This makes it very well suited to formalize state-based machines such as AMMs, where it is crucial to check the correctness of the imposed constraints across sequences of transactions. On the other hand, as we previously explained, some of the mathematical libraries might be less mature compared to other proof assistants and may require additional effort to formalize certain proof strategies. 
While other proof assistants such as Isabelle/HOL~\cite{Nipkow02isabelle} provide strong support for real analysis and automation, the present formalization builds directly on an existing Lean 4 development and follows an operational, state-based modeling style that is well supported in Lean. Reproducing the same results in another proof assistant would have required re-implementing the base model from scratch and might have involved different, and currently unexplored, technical challenges.

\paragraph*{Discussion and Future Work}

We envision the following directions for future research:

\begin{enumerate}
    \item \textbf{Guarded transactions and Slippage Protection}: A natural extension of our model, bringing it closer to the actual implementation of Uniswap v2, would be to formalise the parameters \codelst{amountOutMin} and \codelst{deadline} of the function \codelst{swapExactTokensForTokens} (\Cref{sec:uniswap-compare}), which are not currently represented in our model. 
    The former enforces a minimum acceptable amount of output tokens to the user executing the swap, while the latter prevents transactions from remaining pending for unbounded period, reducing the risk of adverse price movements and slippage. 

    \item \textbf{Concentrated Liquidity and Multi-Fee tiers}: 
    Uniswap v3 \cite{uniswapv3} introduces so-called \emph{concentrated liquidity}: instead of depositing liquidity across the whole range of the internal exchange rate, in Uniswap v3 LPs can specify an internal exchange rate range in which their liquidity earn fees. This allows for different strategies for LPs which can increase their gain but also the risk. Another feature that was introduced in Uniswap v3 is the one of multiple fee tiers. Uniswap v2 has a flat fee of 0.3\% on every pool, while in Uniswap v3 it is possible to decide on four discrete tiers upon the creation of the pool, which are 0.01\%, 0.05\%, 0.3\% and 1\%. Clearly, a higher fee tier benefits the LPs, but could also result in a less attractive pool to trade on for arbitrageurs. Formalizing these features in the current model would bring it closer to recent AMM implementations such as v3. 

    \item \textbf{Incentives for Liquidity Providers}: 
    This work mainly focuses on the arbitrageur's perspective, \ie determining  the optimal swap amount that maximise an arbitrageur's gain. We do not, however, analyse the impact of trading fees on liquidity providers. 
    Namely, while we know that trading fees increase their share of the pool's reserves compared to the fee-less case, a quantitative assessment of this effect and an analysis of trading fees from the liquidity providers' perspective remain open and relevant directions. 
    
    \item \textbf{Lean 4 Automation}: 
    The implementation developed in this work could be significantly simplified by making it more amenable to proof automation.
    As discussed earlier (see ``Challenges of formalizing AMMs in Lean 4'' at page~\pageref{sec:challenges-lean4}), automation is currently hindered by several factors.
    One promising direction would be to replace the subtype of positive real numbers --- which is used throughout the library --- with $\mathbb{R}$, as well as to systematically review the associated simplification lemmas. 
\end{enumerate}

\bibstyle{plainurl} 
\bibliography{main}

\appendix
\section{Comparison with Uniswap v2}
\label{sec:uniswap-compare}

In this section we analyze the differences and similarities between the swap used in this work's model, and the swap used in Uniswap v2's implementation. The Uniswap one is slightly more complicated than the one presented in this model, and the main differences are: 
\begin{enumerate}
    \item Uniswap allows to swap both tokens $\tokT[0]$ and $\tokT[1]$ at the same time, while our model allows only one atomic token type as input. 
    \item In Uniswap, when calling the function, the input is not the amount of tokens that the user wants to swap, but instead the maximum amount of tokens that the user expects to receive. The input token amount is then computed internally in the swap function.
\end{enumerate}

Uniswap v2 is architected in two main modules, the \emph{core} and the \emph{periphery}. The core module only contains the minimal and critical contracts, while the periphery holds all the helper and router contracts that integrate and extend the core functionalities.
It is worth mentioning that, according to Uniswap v2's guide on implementing a swap in a contract\footnote{\url{https://docs.uniswap.org/contracts/v2/guides/smart-contract-integration/trading-from-a-smart-contract}}, it is recommended not to use the \codelst{swap} primitive from the core module, but instead to use the Router\footnote{https://github.com/Uniswap/v2-periphery/blob/master/contracts/UniswapV2Router02.sol} from the periphery module. This is because the Router provides a set of methods that allow users to safely swap the desired tokens. For example, in our model, a swap action \mbox{$\actAmmSwapExact{\pmvA}{}{x}{\tokT[0]}{\tokT[1]}$} means that $\pmvA$ wants to swap exactly an amount $x$ of $\tokT[0]$ tokens, and expects $y = x \cdot \SX{x, \ammR[0], \ammR[1]}$ of $\tokT[1]$ tokens as output. For this scenario, the Router provides a function \codelst{swapExactTokensForTokens} which does exactly what we want. To help the reader understand the similarities and differences from our implementation to the one of Uniswap v2, suppose that we want to perform a swap \mbox{$\actAmmSwapExact{\pmvA}{}{x}{\tokT[0]}{\tokT[1]}$} in a state $\confG = \amm{\ammR[0] : \tokT[0]}{\ammR[1] : \tokT[1]} \mid \Delta$. Then, we explain how this swap is executed using Uniswap Router02's \codelst{swapExactTokensForTokens}, analyzing the variables and function calls being made, up until the swap primitive presented in \Cref{lst:swap}.

\Cref{lst:router-swap-exact} shows \codelst{swapExactTokensForTokens}'s implementation. In particular, we can see that it takes as input: 
\begin{enumerate}
    \item \codelst{amountIn}: this corresponds to our $x$ in the swap

    \item \codelst{amountOutMin}: this specifies the minimum amount of output tokens that the user $\pmvA$ expects.
    In this work's model, we assume that this value is zero, as the rule \nrule{[Swap]} does not express a minimum amount of tokens that the user can expect upon a swap.

    \item \codelst{path}: this is the array of token addresses that the trade goes through. The first of these is the input token, and the last one is the output token. So, in our case this is going to be simply $[\tokT[0], \tokT[1]]$.

    \item \codelst{to}: this is the address that receives the final output tokens. In our case, this address coincides with $\pmvA$, the user performing the swap.

    \item \codelst{deadline}: this is a timestamp after which the router reverts the swap if it has not been included in a block yet. This is mainly to prevent bad exchange rates if the transaction takes too long to be approved.
    Since the current model does not represent the transaction mempool, this feature is omitted. 
\end{enumerate}

\begin{figure}
\vbox{%
\small
\begin{lstlisting}[caption={Method \codelst{swapExactTokensForTokens} in Router02.},label={lst:router-swap-exact}]
function swapExactTokensForTokens(
        uint amountIn,
        uint amountOutMin,
        address[] calldata path,
        address to,
        uint deadline
    ) external virtual override ensure(deadline) returns (uint[] memory amounts) {
        amounts = UniswapV2Library.getAmountsOut(factory, amountIn, path);
        require(amounts[amounts.length - 1] >= amountOutMin, 'UniswapV2Router: INSUFFICIENT_OUTPUT_AMOUNT');
        TransferHelper.safeTransferFrom(
            path[0], msg.sender, UniswapV2Library.pairFor(factory, path[0], path[1]), amounts[0]
        );
        _swap(amounts, path, to);
}
\end{lstlisting}}
\end{figure}

To better understand what \codelst{swapExactTokensForTokens} does in its body, we first explain how the output amount $y$ is computed, then break down how the transfer of the input tokens amount $x$ happens, and finally analyze the swap primitive call.

\subsection{Output amount}
In line 8, the function calls the library function \codelst{getAmountsOut} \footnote{\url{https://github.com/Uniswap/v2-periphery/blob/master/contracts/libraries/UniswapV2Library.sol\#L62-L70}}, the code of which is reported in ~\Cref{lst:lib-amounts-out}. This function computes an array of amounts of output tokens depending on how many there are specified in the parameter \codelst{path}, where the first element of the array is the input token amount \codelst{amountIn}. In our case, since we are trading in the pool $\amm{\ammR[0] : \tokT[0]}{\ammR[1] : \tokT[1]}$ and we already explained that \codelst{path} $= [\tokT[0], \tokT[1]]$, we will have \codelst{amounts[0]} $=$ \codelst{amountIn} and \codelst{amounts[1]} $=$ \codelst{getAmountOut(amounts[0], reserveIn, reserveOut)} where \codelst{reserveIn} and \codelst{reserveOut} are exactly our $\ammR[0]$ and $\ammR[1]$.

The function \codelst{getAmountOut} \footnote{\url{https://github.com/Uniswap/v2-periphery/blob/master/contracts/libraries/UniswapV2Library.sol\#L43-L50}}(\Cref{lst:lib-amount-out}) is where the constant product swap rate is implemented. In comparison with our model, this returns exactly an amount $y = x \cdot \SX{x, \ammR[0], \ammR[1]}$ of $\tokT[1]$ tokens. Analyzing the code line-by-line, the variables being computed are \codelst{amountInWithFee} $ = x \cdot 997$, \codelst{numerator} $ = x \cdot 997 \cdot \ammR[1]$,  \codelst{denominator} $ = \ammR[0] \cdot 1000 + x \cdot 997$ and \codelst{amountOut} $= y$. Putting all together: 
\begin{align*}
    y & = \frac{\text{\codelst{numerator}}}{\text{\codelst{denominator}}} = \frac{x \cdot 997 \cdot \ammR[1]}{\ammR[0] \cdot 1000 + x \cdot 997}
    = 
    \frac{1000 \cdot (x \cdot 0.997 \cdot \ammR[1])}{1000 \cdot (\ammR[0] + 0.997 \cdot x)}
    = \frac{x \cdot 0.997 \cdot \ammR[1]}{\ammR[0] + 0.997 \cdot x}
\end{align*}

which is exactly the constant product swap rate from Definition~\ref{defi:const-prod} with $\fee = 0.997$, the trading fee hard-coded in Uniswap v2. 

Going back to line 8 of ~\Cref{lst:router-swap-exact}, \codelst{amounts} will have the input amount $x$ for the first element, and the output amount $y = x \cdot \SX{x, \ammR[0], \ammR[1]}$ that we just computed as the second one. In line 9, this safety check ensures that the output amount $y$ is at least what the user $\pmvA$ expects to receive as a minimum amount of output tokens. If not, the transaction is reverted. 

\subsection{Tokens transfer}

In lines 10-11, we see the transfer of $x$ $\tokT[0]$ tokens into the contract of the pair $\tokM{\tokT[0]}{\tokT[1]}$ before calling the function \codelst{\_swap}. This is done because the \codelst{\_swap} function expects the input tokens to be already transferred into the pair. The code of this function is in \Cref{lst:router-swap}. This is the function that prepares the parameters to call the \codelst{swap} primitive, illustrated in \Cref{lst:swap}. In particular, we can see that in line 6 the function instantiates \codelst{amount0Out} and \codelst{amount1Out}, which will be the parameters passed to the \codelst{swap} primitive. In this case, depending on the ordering of the tokens $\tokT[0]$ and $\tokT[1]$ in the AMM pair, one of \codelst{amount0Out} and \codelst{amount1Out} will be instantiated with our output amount $y$, and the other one to zero, since we do not expect two output amounts, but only one of $\tokT[1]$ tokens. For simplicity, suppose from now on that in our case \codelst{amount0Out} $= 0$, since we are swapping $\tokT[0]$ tokens. 

\begin{figure}
\vbox{%
\small
\begin{lstlisting}[caption={Method \codelst{getAmountsOut}.},label={lst:lib-amounts-out}]
function getAmountsOut(address factory, uint amountIn, address[] memory path) internal view returns (uint[] memory amounts) {
        require(path.length >= 2, 'UniswapV2Library: INVALID_PATH');
        amounts = new uint[](path.length);
        amounts[0] = amountIn;
        for (uint i; i < path.length - 1; i++) {
            (uint reserveIn, uint reserveOut) = getReserves(factory, path[i], path[i + 1]);
            amounts[i + 1] = getAmountOut(amounts[i], reserveIn, reserveOut);
        }
    }
\end{lstlisting}}
\end{figure}

\begin{figure}
\vbox{%
\small
\begin{lstlisting}[caption={Method \codelst{getAmountOut}.},label={lst:lib-amount-out}]
 function getAmountOut(uint amountIn, uint reserveIn, uint reserveOut) internal pure returns (uint amountOut) {
        require(amountIn > 0, 'UniswapV2Library: INSUFFICIENT_INPUT_AMOUNT');
        require(reserveIn > 0 && reserveOut > 0, 'UniswapV2Library: INSUFFICIENT_LIQUIDITY');
        uint amountInWithFee = amountIn.mul(997);
        uint numerator = amountInWithFee.mul(reserveOut);
        uint denominator = reserveIn.mul(1000).add(amountInWithFee);
        amountOut = numerator / denominator;
    }
\end{lstlisting}}
\end{figure}

Next, since we are dealing with only one token pair, when the function computes \codelst{to}, it is immediately set to \codelst{_to} (which in our case is $\pmvA$'s wallet). The final line represents the call to the \codelst{swap} primitive, where 

\begin{enumerate}
    \item \codelst{amount0Out} is zero, as already explained, since we are swapping $\tokT[0]$ tokens, and we don't expect to receive any $\tokT[0]$ tokens as output. 

    \item \codelst{amount1Out} is equal to $y = \SX{x, \ammR[0], \ammR[1]}$.

    \item \codelst{to} represents $\pmvA$'s wallet.
\end{enumerate}

\subsection{Swap primitive and Invariant check}

Finally, we investigate the function \codelst{swap} in \Cref{lst:swap}. As previously stated, this function is slightly complicated, so we only want to give an intuition on its behavior for our example case. We begin by locating where our input and output amounts $x$ and $y$ are located/computed inside the function: 

\begin{figure}
\vbox{%
\small
\begin{lstlisting}[caption={Method \codelst{_swap} in Router02.},label={lst:router-swap}]
function _swap(uint[] memory amounts, address[] memory path, address _to) internal virtual {
        for (uint i; i < path.length - 1; i++) {
            (address input, address output) = (path[i], path[i + 1]);
            (address token0,) = UniswapV2Library.sortTokens(input, output);
            uint amountOut = amounts[i + 1];
            (uint amount0Out, uint amount1Out) = input == token0 ? (uint(0), amountOut) : (amountOut, uint(0));
            address to = i < path.length - 2 ? UniswapV2Library.pairFor(factory, output, path[i + 2]) : _to;
            IUniswapV2Pair(UniswapV2Library.pairFor(factory, input, output)).swap(
                amount0Out, amount1Out, to, new bytes(0)
            );
        }
    }
\end{lstlisting}}
\end{figure}

\begin{enumerate}
    \item The input amount $x$ is computed inside the function in line 18, called \codelst{amount0In}. Remember that inside the Router's \codelst{swapExactTokensForTokens} we have already performed a safe transfer of this input amount $x$. Hence, the function can simply compute \codelst{amount0In} as new reserves of $\tokT[0]$ called \codelst{balance0} minus the old reserves, called \codelst{\_reserve0}. Intuitively, since we did not transfer any $\tokT[1]$ token to the pair, then \codelst{amount1In} is zero, as expected. 

    \item The output amount $y$ is passed as the parameter \codelst{amount1Out} like previously said. Also, remember that the other parameter \codelst{amount0Out} is zero, from the computations made in the function \codelst{_swap}. 
\end{enumerate}

There is only one thing left to check: the invariant. So far, we have not found anything that checks that $(\ammR[0]+x) (\ammR[1]-y) \; \geq \; \ammR[0] \ammR[1]$. This is done in an even stricter way inside the \codelst{swap} function. 
In line 22-23 the function declares \codelst{balance0Adjusted} and \codelst{balance1Adjusted}, which are computed as: 
\begin{align*}
& \text{\codelst{balance0Adjusted}} = (\ammR[0] + x)\cdot 1000 - 3x
\\
& \text{\codelst{balance1Adjusted}} =(\ammR[1] -y)\cdot1000 - 3\cdot0 = (\ammR[1] - y)\cdot1000
\end{align*}
Recall that at this point of the contract, \codelst{balance0} and \codelst{balance1} are respectively $\ammR[0]$ and $\ammR[1]$ after the transfers have occurred, \ie $\ammR[0] + x$ and $\ammR[1] - y$. Then in line 24, the \codelst{require} specifies the invariant check:
    \[\text{\codelst{balance0Adjusted}} \cdot \text{\codelst{balance1Adjusted}} \geq \ammR[0]\ammR[1]\cdot1000^2\]
    At first sight this check might seem very different from the one of our model. However, we can show that it is almost identical: 
    \begin{align*}
        & \text{\codelst{balance0Adjusted}} \cdot \text{\codelst{balance1Adjusted}} \geq \ammR[0]\ammR[1]\cdot1000^2
        \\ \iff &  
        ((\ammR[0] + x)\cdot 1000 - 3x) ((\ammR[1] - y)\cdot1000) \geq \ammR[0]\ammR[1]\cdot1000^2
        \\ \iff &  
        (\ammR[0] + x - 0.003x) (\ammR[1] - y) \geq \ammR[0]\ammR[1]
        \\ \iff &  
        (\ammR[0] + 0.997x) (\ammR[1] - y) \geq \ammR[0]\ammR[1]
        \\ \iff &  
        (\ammR[0] + \fee x) (\ammR[1] - y) \geq \ammR[0]\ammR[1] \quad && \text{with } \fee = 0.997
    \end{align*}
    This check is even more strict than the one shown in Section~\ref{sec:amm-model}, as it only takes into account the fee-reduced input $\fee x$. This check clearly implies the previous one, as $\ammR[0] + \fee x < \ammR[0] + x$.

\begin{figure}
\vbox{%
\small
\begin{lstlisting}[caption={Uniswap v2's method \codelst{swap}.},label={lst:swap}]
function swap(uint amount0Out, uint amount1Out, address to, bytes calldata data) external lock {
    require(amount0Out > 0 || amount1Out > 0, 'UniswapV2: INSUFFICIENT_OUTPUT_AMOUNT');
    (uint112 _reserve0, uint112 _reserve1,) = getReserves(); // gas savings
    require(amount0Out < _reserve0 && amount1Out < _reserve1, 'UniswapV2: INSUFFICIENT_LIQUIDITY');

    uint balance0;
    uint balance1;
    { // scope for _token{0,1}, avoids stack too deep errors
    address _token0 = token0;
    address _token1 = token1;
    require(to != _token0 && to != _token1, 'UniswapV2: INVALID_TO');
    if (amount0Out > 0) _safeTransfer(_token0, to, amount0Out); // optimistically transfer tokens
    if (amount1Out > 0) _safeTransfer(_token1, to, amount1Out); // optimistically transfer tokens
    if (data.length > 0) IUniswapV2Callee(to).uniswapV2Call(msg.sender, amount0Out, amount1Out, data);
    balance0 = IERC20(_token0).balanceOf(address(this));
    balance1 = IERC20(_token1).balanceOf(address(this));
    }
    uint amount0In = balance0 > _reserve0 - amount0Out ? balance0 - (_reserve0 - amount0Out) : 0;
    uint amount1In = balance1 > _reserve1 - amount1Out ? balance1 - (_reserve1 - amount1Out) : 0;
    require(amount0In > 0 || amount1In > 0, 'UniswapV2: INSUFFICIENT_INPUT_AMOUNT');
    { // scope for reserve{0,1}Adjusted, avoids stack too deep errors
    uint balance0Adjusted = balance0.mul(1000).sub(amount0In.mul(3));
    uint balance1Adjusted = balance1.mul(1000).sub(amount1In.mul(3));
    require(balance0Adjusted.mul(balance1Adjusted) >= uint(_reserve0).mul(_reserve1).mul(1000**2), 'UniswapV2: K');
    }

    _update(balance0, balance1, _reserve0, _reserve1);
    emit Swap(msg.sender, amount0In, amount1In, amount0Out, amount1Out, to);
}
\end{lstlisting}}
\end{figure}

\end{document}